\documentclass[journal]{IEEEtran}

\usepackage{cite}
\usepackage{amsmath}
\usepackage[T1]{fontenc}
\usepackage[latin9]{inputenc}
\usepackage{units}
\usepackage{textcomp}
\usepackage{stmaryrd}
\usepackage[caption=false,font=footnotesize]{subfig}
\usepackage{graphicx}
\usepackage[english]{babel}
\usepackage{stfloats}
\usepackage{fixltx2e}
\usepackage{array}
\usepackage{color}
\usepackage{url}
\makeatletter

\providecommand{\tabularnewline}{\\}

\makeatother

\begin{document}

\title{Selective Light Field Refocusing for Camera Arrays Using Bokeh Rendering and Superresolution}

\author{Yingqian~Wang,~~Jungang~Yang,~~Yulan~Guo,~~Chao~Xiao,~~Wei~An

\thanks{This work was partially supported by the National Natural Science Foun-dation of China (Nos. 61401474, 61602499 and 61471371), the Hunan Pro-vincial National Science Foundation (No. 2016JJ3025), the National Postdoc-toral Program for Innovative Talents (No. BX201600172), and the China Postdoctoral Science Foundation. (Corresponding author: Jungang Yang) }

\thanks{Y.~Wang, J.~Yang, Y.~Guo, C.~Xiao, and W.~An are with the College of Electronic Science and Technology, National University of Defense Technology, Changsha, China. (e-mail: wangyingqian16@nudt.edu.cn, yangjungang@nudt.edu.cn, yulan.guo@nudt.edu.cn, xiaochao12@nudt.edu.cn, anwei@nudt.edu.cn). Y.~Guo is also with the School of Electronics and Communication Engineering, Sun Yat-sen University, Guangzhou, China.}}

\markboth{IEEE Signal Processing Letters}%
{Shell \MakeLowercase{\textit{et al.}}: Bare Demo of IEEEtran.cls for IEEE Journals}
\maketitle

\begin{abstract}
Camera arrays provide spatial and angular information within a single snapshot. With refocusing methods, focal planes can be altered after exposure. In this letter, we propose a light field refocusing method to improve the imaging quality of camera arrays. In our method, the disparity is first estimated. Then, the unfocused region (bokeh) is rendered by using a depth-based anisotropic filter. Finally, the refocused image is produced by a reconstruction-based super-resolution approach where the bokeh image is used as a regularization term. Our method can selectively refocus images with focused region being super-resolved and bokeh being aesthetically rendered. Our method also enables post-adjustment of depth of field. We conduct experiments on both public and self-developed datasets. Our method achieves superior visual performance with acceptable computational cost as compared to other state-of-the-art methods. Code is available at \url{https://github.com/YingqianWang/Selective-LF-Refocusing}.
\end{abstract}


\section{Introduction}

\IEEEPARstart{A}{lthough} miniature cameras (e.g., smartphone cameras) have become increasingly popular in recent years, professional photographers still prefer traditional digital single lens reflex (DSLR) to generate aesthetical photographs. Due to the large-size image sensors and large-aperture lens, images produced by DSLRs are high in resolution, high in signal-to-noise ratio (SNR), and can be shallow in depth of field (DoF). That is, only a limited range of depths are in focus, leaving objects in other depths suffering from varying degrees of blurs (termed as \textit{bokeh}). Using bokeh effect properly can blur out non-essential elements in foreground or background, and guide the visual attention of audiences to major objects in a photograph.

However, most professional DSLRs tend to be bulk and expensive due to their high-quality lenses and sensors. Recently, with the rapid development of light field (LF) imaging \cite{kanojia2017postcapture, yoon2017light, yuan2018light},
it becomes possible to use low-cost cameras to take photographs with a high quality as DSLRs. Several designs \cite{lin2014separable, ng2005light, georgiev2010focused, wilburn2005high, vaish2006reconstructing, vaish2004using, venkataraman2013picam, lin2015camera} have been proposed to capture LF. Among them, plenoptic cameras \cite{ng2005light} place an array of microlens between main lens and image sensors to capture the angular information of a scene. However, due to the limitation of sensor resolution, one can only opt for a dense angular sampling \cite{ng2005light}, or a dense spatial sampling \cite{georgiev2010focused}. In contrast, camera arrays \cite{wilburn2005high, vaish2006reconstructing, vaish2004using, venkataraman2013picam} set several independent cameras in a regular grid, and therefore have an improved image resolution. Wu \textit{et al}. \cite{Wu2017Light} claim that although early camera array systems \cite{wilburn2005high, vaish2006reconstructing, vaish2004using} are bulk and hardware-intensive, these LF acquisition systems have a bright prospective with camera miniaturization being exploited and small-size camera arrays \cite{venkataraman2013picam,lin2015camera} being developed. Compared to plenoptic cameras, camera arrays have wider baselines and sparser angular samplings. Consequently, traditional refocusing methods \cite{vaish2006reconstructing, vaish2004using, levoy2004synthetic} face aliasing problems in the bokeh regions.

LF reconstruction methods \cite{yoon2017light, farrugia2017super, Wanner2014Variational} are widely used to address the aliasing problem. These methods interpolate novel views between existing views, and use the synthetic LF to refocus sub-images. Specifically, methods \cite{yoon2017light, farrugia2017super, Wanner2014Variational} can simultaneously increase the number of viewpoints and image resolution. Since there is a trade-off between the density of views and computational cost, methods \cite{yoon2017light, farrugia2017super, Wanner2014Variational} either suffer from a high computational burden, or remain aliasing artifacts to some degree. Bokeh rendering methods \cite{potmesil1981lens,lee2008real,lee2009real,bertalmio2004real,liu2016stereo} provide novel paradigms. Methods \cite{potmesil1981lens,lee2008real} map source pixels onto circles of confusion (CoC) \cite{potmesil1981lens} using the estimated depth information, and blend CoC in the order of depth. Although \cite{potmesil1981lens, lee2008real} can eliminate aliasing artifact, the required depth sorting process is costly . Methods \cite{lee2009real, bertalmio2004real, liu2016stereo} filter images according to the CoC size, and achieve a compromise between computational cost and bokeh performance. In summary, methods \cite{potmesil1981lens, lee2008real, lee2009real, bertalmio2004real, liu2016stereo} can improve bokeh performance, but cannot increase image resolution.

In this paper, we propose an LF refocusing method to improve the imaging quality of camera arrays. The outline of our method is illustrated in Fig. 1. The main contributions of this paper can be summarized as follows: (1) Bokeh rendering and super-resolution (SR) are selectively integrated into one scheme, i.e., our method can simultaneously improve bokeh performance and increase image resolution. (2) DoF can be adjusted in a post-processing mode, which is important for producing aesthetical photographs on different scenes, and it cannot be achieved by DSLRs or previous refocusing methods. (3) We conducted experiments on a self-developed dataset. Images generated by our method are more similar to those generated by a DSLR than existing methods.

\begin{figure}

\centering\includegraphics[width=8.5cm]{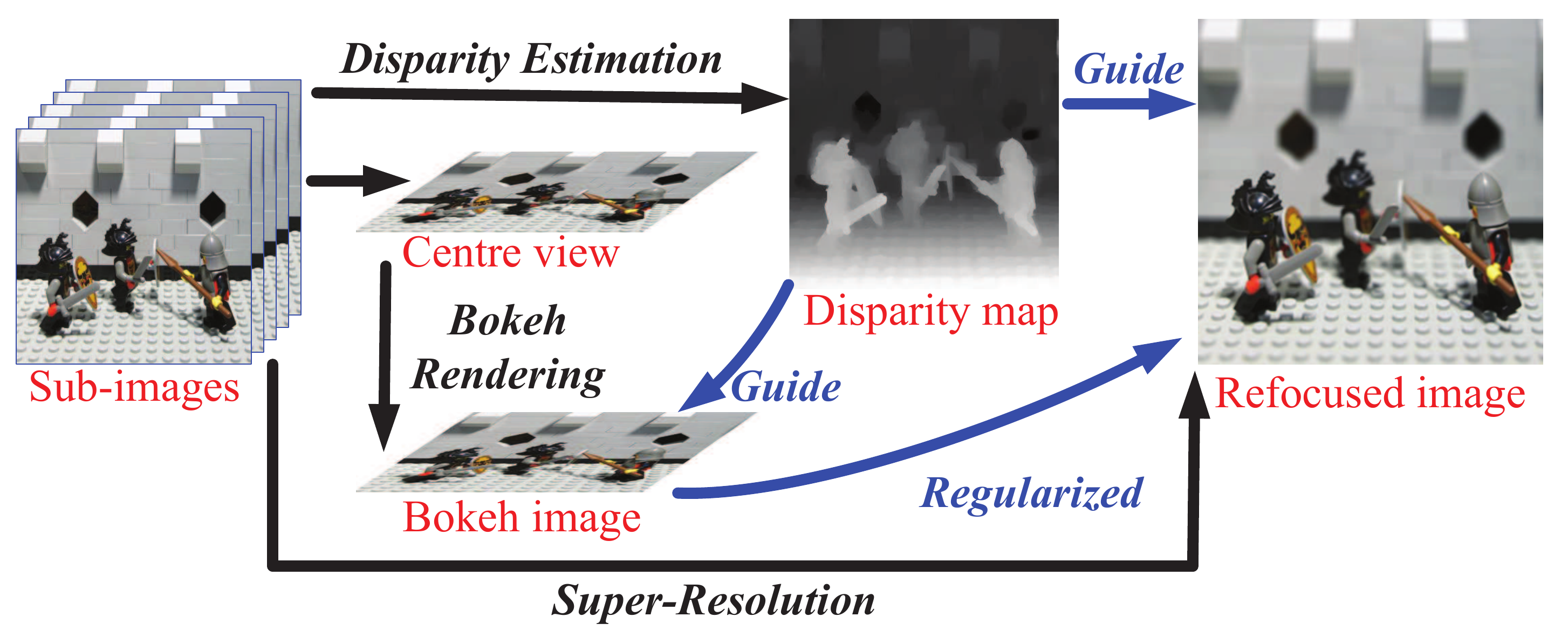}
\caption{Outline of our method. The disparity map is first estimated. The bokeh is then rendered on the extracted center-view sub-image. Finally, the refocused image is generated by a reconstruction-based SR approach where the rendered bokeh image is used as a regularization term.}
\end{figure}

\section{Reconstruction-based SR model}

Sub-images captured by camera arrays suffer from shearing shifts (caused by parallax), blurring (caused by optical distortions), and down-sampling (caused by low-resolution image sensors). Considering these factors, we formulate the degradation model \cite{farsiu2004fast} of a camera array as
\begin{equation}
\boldsymbol{y}_{k}=\boldsymbol{D}\boldsymbol{H}\boldsymbol{F}_{i,k}\boldsymbol{x}+\boldsymbol{n}_{k},
\end{equation}
where $\boldsymbol{x}$, $\boldsymbol{y}_{k}$, and $\boldsymbol{n}_{k}$ denote high-resolution (HR) image, sub-image captured by the $k^{th}$ camera, and noise of the $k^{th}$ sub-image, respectively. $N$ represents the number of cameras in the array, and $n$ represents depth resolution (i.e., total number of depth layers). $\boldsymbol{H}$,
$\boldsymbol{D}$, and $\boldsymbol{F}_{i,k}$ denote optical blurring, down-sampling, and shifts (dependent of depth $i$ and viewpoint $k$), respectively. Since the main task of SR reconstruction is to estimate $\boldsymbol{x}$ to suit the degradation model in Eq. (1), we formulate the SR model as a task to minimize the following function:
\begin{align}
\hat{\boldsymbol{x}} & =\underset{\boldsymbol{x}}{\arg\min}\{\sum_{k=1}^{N}\left\Vert \left(\boldsymbol{1}-\boldsymbol{\omega}_{b}\right)\varodot\left(\boldsymbol{y}_{k}-\boldsymbol{D}\boldsymbol{H}\boldsymbol{F}_{i,k}\boldsymbol{x}\right)\right\Vert _{2}^{2}\\
 & +\lambda_{b}J_{b}\left(\boldsymbol{x}\right)+\lambda_\textit{BTV}J_\textit{BTV}\left(\boldsymbol{x}\right)\},\qquad i=1,2,\cdots,n\nonumber.
\end{align}

The first term in Eq. (2) denotes the distance between observations and an ideal HR image. $J_{b}\left(\boldsymbol{x}\right)$ is the bokeh regularization
term and $J_\textit{BTV}\left(\boldsymbol{x}\right)$ is the bilateral total variation (BTV) regularization term. Readers are referred to \cite{farsiu2004fast} for more details of BTV regularization. $\lambda_{b}$ and $\lambda_\textit{BTV}$ are regularization weights. $\boldsymbol{1}$ denotes a vector with all elements equal to 1, and $\varodot$ denotes element-wise multiplication. $\boldsymbol{\omega}_{b}$ is a depth-based and spatial-variant weight vector where unfocused regions share larger values. $J_{b}\left(\boldsymbol{x}\right)$ can be expressed as
\begin{equation}
J_{b}\left(\boldsymbol{x}\right)=\left\Vert \boldsymbol{\omega}_{b}\varodot\left(\boldsymbol{x}-\boldsymbol{x}_{b}\right)\right\Vert _{2}^{2},
\end{equation}
where $\boldsymbol{x}_{b}$ represents the bokeh image. More details
on $\boldsymbol{x}_{b}$ and $\boldsymbol{\omega}_{b}$ will be presented
in Section III.
We use the gradient descend approach \cite{farsiu2004fast} to
approximate the optimal solution. The settings of step size and number of iterations (NoI) will be introduced in Section~IV. Readers can refer to \cite{wang2018fast} for the convergence issue of SR process.

\section{Bokeh Rendering}

As described in Section~II, the key step of our method is to generate
the bokeh image $\boldsymbol{x}_{b}$. Assuming that point $p$ is out
of focus and corresponds to a CoC in the image, the radius of CoC
can be calculated as
\begin{equation}
r=\left|\frac{f^{2}\left(\gamma_{f}-\gamma_{p}\right)}{2F\gamma_{p}\left(\gamma_{f}-f\right)}\right|,
\end{equation}
where $\gamma_{f}$ and $\gamma_{p}$ represent the depth of focus
and the depth of $p$, respectively. $f$ is the focal length, $F$
is the F-number of the lens. From the LF model presented in \cite{wang2018disparity},
the depth can be expressed as $\gamma=fB/d$, where $B$ is the length
of the baseline, and $d$ represents the disparity. Consequently,
we can rewrite Eq. (4) as
\begin{equation}
r=\left|\frac{f\left(d_{p}-d_{f}\right)}{2F\left(B-d_{f}\right)}\right|=K\left|d_{p}-d_{f}\right|.
\end{equation}

Note that, the radius of CoC is proportion to the absolute disparity
difference between point $p$ and the focus. Since $f$, $F$, $B$,
and $d_{f}$ are constant during a bokeh rendering process, we consider
$K=f/2F\left(B-d_{f}\right)$ as \textit{bokeh intensity} which represents
the overall degree of bokeh and reflects DoF. A large $K$ corresponds
to a strong bokeh and accordingly, a shallow DoF.

\begin{figure}

\centering
\includegraphics[width=3cm]{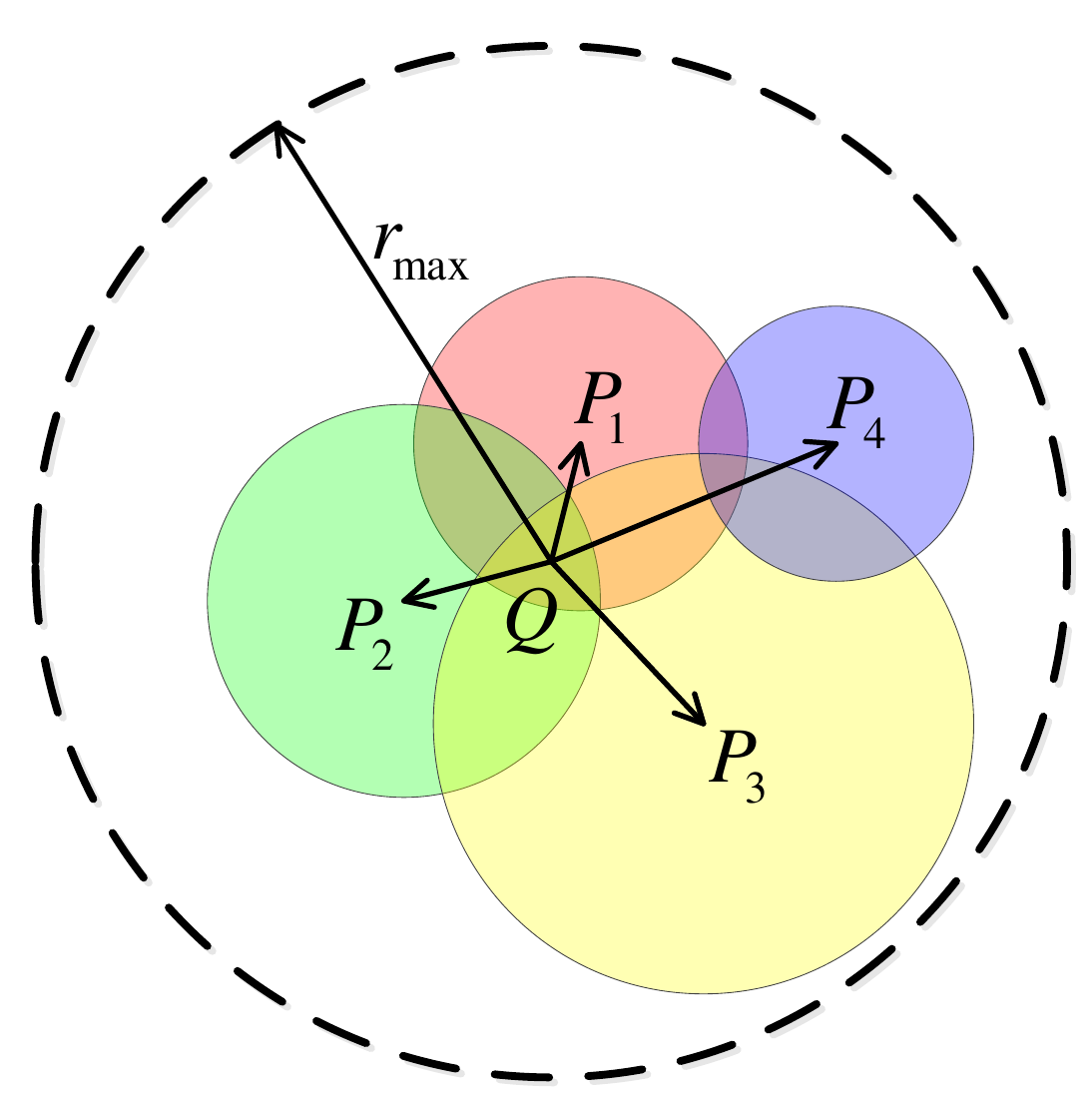}
\caption{An illustration of the bokeh rendering process. To calculate the intensity at Q, contributions from the surrounding CoC have to be combined.}

\end{figure}

We attribute bokeh rendering to anisotropic filtering. As shown in
Fig. 2, $Q$ is surrounded by four CoC centered at $P_{1}$ to $P_{4}$.
Since intensities in a CoC are uniformly distributed,
the contribution from $P_{i}$ to $Q$ can be calculated as $I_{P_{i}\rightarrow Q}=\nicefrac{I_{P_{_{i}}}}{\pi r_{P_{i}}^{2}},\:\left(\varphi_{P_{i}Q}\leq r_{P_{i}}\right)$,
where $r_{P_{i}}$ is the radius of CoC centered at $P_{i}$, $I_{P_{i}}$
is the intensity at $P_{i}$ before bokeh rendering, $\varphi_{P_{i}Q}$
represents the distance between $P_{i}$ and $Q$. Note that, $P_{i}$
has an impact on $Q$ only if $Q$ is within the CoC of $P_{i}$,
i.e., $\varphi_{P_{i}Q}\leq r_{P_{i}}$. To calculate the intensity
at $Q$, the contributions of surrounding points have to be combined:
\begin{equation}
I_{Q}=\sum_{P_{i}\in\varOmega_{Q}}\omega_{P_{i}}I_{P_{i}\rightarrow Q},
\end{equation}
where $\varOmega_{Q}=\left\{ P_{i}|\varphi_{P_{i}Q}\leq r_{max}\right\} $
represents the set of points around $Q$, $r_{max}$ is the maximum
radius of CoC in the image. Note that, some points in $\varOmega_{Q}$
may have no contribution to $Q$, e.g., point $P_{4}$ in Fig. 2.
Taken this into consideration, weight $\omega_{P_{i}}$ is calculated as
\begin{equation}
\omega_{P_{i}}=\begin{cases}
\nicefrac{1}{\pi r_{P_{i}}^{2}}, & \varphi_{P_{i}Q}\leq r_{P_{i}}\\
0, & r_{P_{i}}\leq\varphi_{P_{i}Q}\leq r_{max}
\end{cases}.
\end{equation}

An anisotropic filter can be used to render the extracted center view based on Eqs. (6) and (7). By using bicubic interpolation, the bokeh image $\boldsymbol{x}_{b}$ is generated by up-sampling the rendered center view to the target resolution.

Another significant step is to calculate $\boldsymbol{\omega}_{b}$
used in Eqs. (2) and (3). Since the blurring degree in an image is determined
by the radius of CoC, we calculate $\boldsymbol{\omega}_{b}$ in two
steps.

Step 1: Normalize the radius to interval $\left[0,1\right]$
by $\eta_{p}=\nicefrac{\left(r_{p}-r_{min}\right)}{\left(r_{max}-r_{min}\right)}$,
where $r_{min}$ is the minimum radius of CoC in the image.

Step 2: To divide $\eta_{p}$ into two parts (focus and bokeh) appropriately,
we use sigmoid function to transform $\eta_{p}$
into $\omega_{p}$ according to $\omega_{p}=\nicefrac{1}{\left(1+exp\left\{ -a\left(\eta_{p}-b\right)\right\} \right)}$,
where $a$ is a decaying factor, and $b$ is the threshold. Finally, $\boldsymbol{\omega}_{b}$
can be obtained by traversing all pixels and re-ordering $\omega_{p}$
into a vector.

\section{Experiments}

Extensive experiments are conducted on both public and self-developed LF
datasets. The 3$\times$3 sub-images of the Stanford LF dataset \cite{vaish2008new} were first down-sampled by a factor of 2 and then used to test different methods. We compared our method with a refocusing method \cite{vaish2004using}, an LF reconstruction method \cite{farrugia2017super},
and a bokeh rendering method \cite{liu2016stereo}.
For method \cite{farrugia2017super}, the 3$\times$3 LF input was angularly rendered to 5$\times$5 views and spatially super-resolved to the original resolution.
To achieve fair comparison between method \cite{liu2016stereo} and our method,
we used the disparity estimation approach proposed in \cite{wang2018disparity}.
We consider the HR center-view sub-image as the groundtruth and quantitatively evaluate the SR performance by calculating PSNRs of focused regions between the results of each method and the groundtruth.
Since there is no groundtruth for the bokeh, we follow the existing bokeh rendering methods \cite{potmesil1981lens,lee2008real,lee2009real,bertalmio2004real,liu2016stereo},
and use visual inspection for evaluation.

For the self-developed LF dataset, we adopted a scanning scheme since view-by-view scanning in static occasion is equivalent to a single shot by a camera array \cite{Wu2017Light}.
We installed an \textit{iPhone 6S} camera (with a $4.8mm \times 3.6mm$ image sensor
and an $F=2.2$, $f=29mm$ lens) on a gantry. Similar to
the approach proposed in \cite{yang2000light}, we shifted the camera to
9 intersections of a $2\times2$ grid (the size of each grid is $5\times5mm^{2}$), and scanned scenes in our laboratory.
The captured images are calibrated by using the method in \cite{zhang2000flexible}.
We then used a \textit{Canon EOS 5D Mark IV} DSLR (with a $36mm\times24mm$ image sensor and an $F=2.2$, $f=50mm$ lens) to provide \textquotedblleft groundtruth\textquotedblright~images.

The parameter setting of our method is listed in Table~I. All the listed parameters are tuned through experiments to achieve good performance. Specifically, the performance of our method can be improved gradually as NoI increases. To achieve a compromise between visual performance and computational cost, we set NoI to 10 in our implementation. Additionally, we adopted the same settings of $J_\textit{BTV}\left(\boldsymbol{x}\right)$ as in \cite{farsiu2004fast}. We observed through experiments that the final result was insensitive to these parameters except $b$, since $b$ has a dominant influence on the classification of focus and bokeh.

All algorithms were implemented in MATLAB on a PC with a 2.40 GHz CPU (Intel Core i7-5500U) and a 12 GB RAM.

\subsection{Visual Performance Comparison}
\begin{figure}[t]
\vspace{-0.3cm}
\centering
\subfloat[]{
\includegraphics[width=2.1cm]{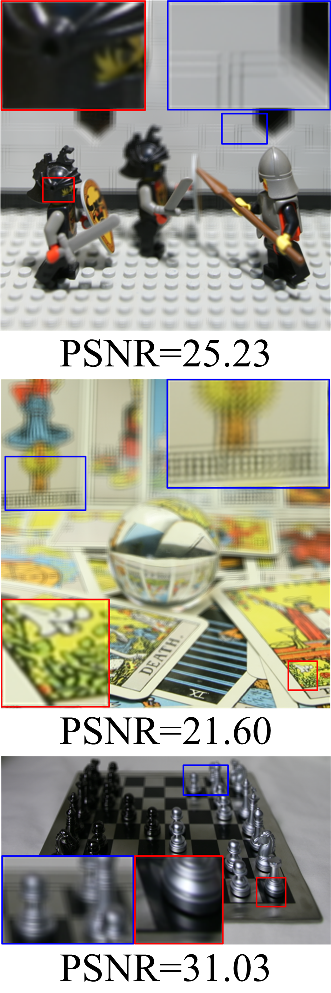}}\subfloat[]{
\includegraphics[width=2.1cm]{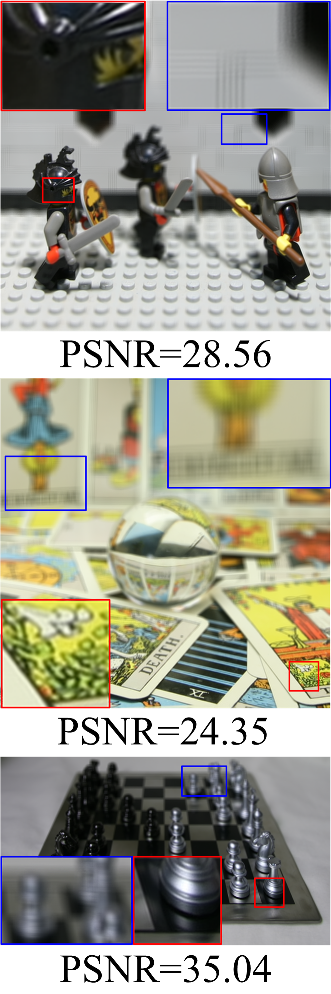}}\subfloat[]{
\includegraphics[width=2.1cm]{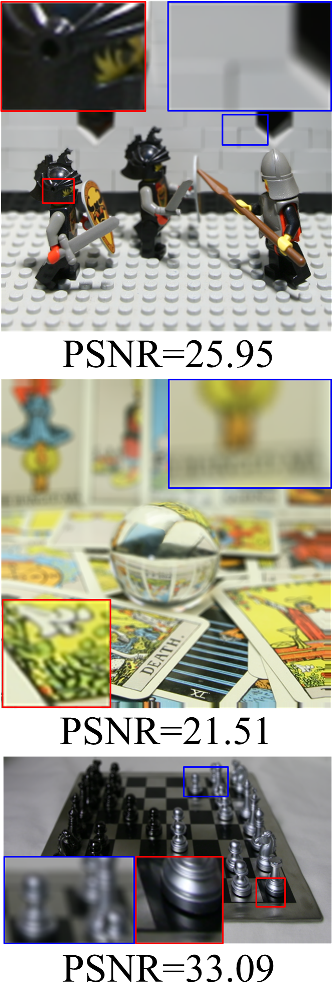}}\subfloat[]{
\includegraphics[width=2.1cm]{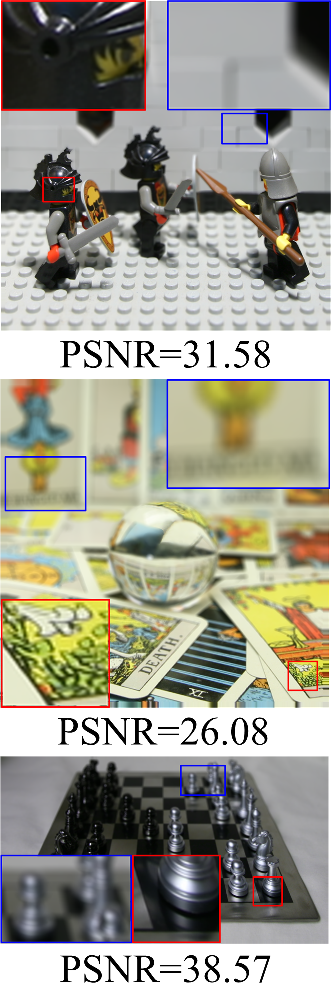}}
\caption{Results achieved on scenes \textit{Knights} (top),
\textit{Cards} (middle), and \textit{Chess} (bottom) in the Stanford LF dataset \cite{vaish2008new}. (a) Vaish \textit{et al}. \cite{vaish2004using}, (b) Farrugia \textit{et al}. \cite{farrugia2017super}, (c) Liu \textit{et al}. \cite{liu2016stereo}, (d) ours ($K$=2). Note that, sharpness of the focused region (images in red boxes) is evaluated quantitatively using PSNR.}

\end{figure}

\begin{table}

\centering
\caption{Parameter Setting}
\vspace{-0cm}
\begin{tabular}{|>{\centering}m{1.5cm}|>{\centering}m{0.5cm}|>{\centering}m{0.8cm}|>{\centering}m{1.2cm}|
>{\centering}m{0.5cm}|>{\centering}m{0.5cm}|>{\centering}m{0.5cm}|}
\hline
Parameters & $\lambda_{b}$ & $\lambda_{BTV}$ & Step Size & NoI & $a$ & $b$\tabularnewline
\hline
Values & 5 & 0.2 & 0.1 & 10 & 15 & 0.3\tabularnewline
\hline
\end{tabular}

\end{table}

\begin{figure*}[t]
\vspace{-0cm}
\centering
\subfloat[Center view]{
\includegraphics[width=2.9cm]{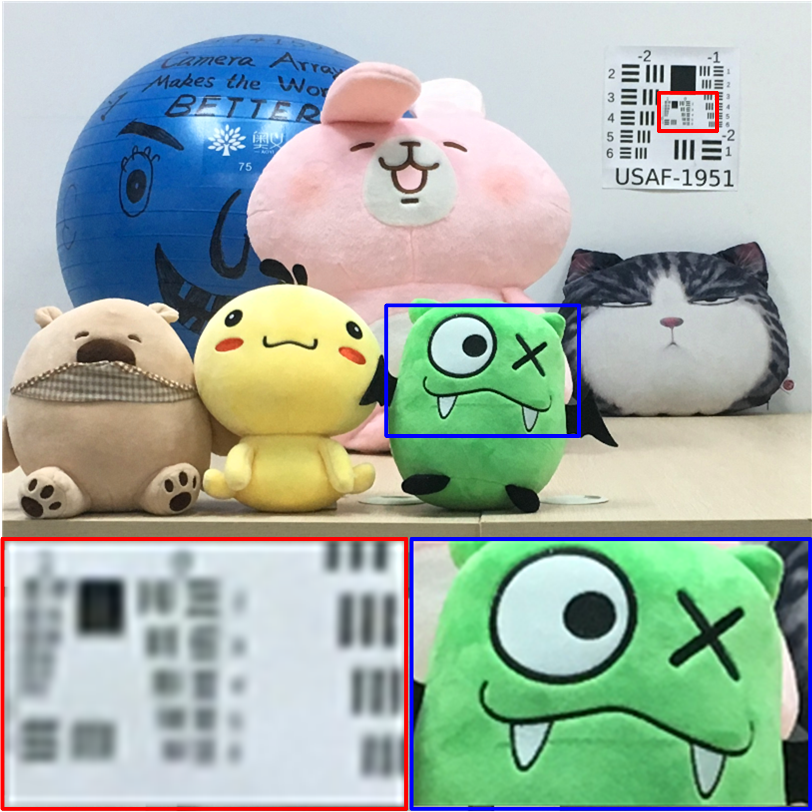}}\subfloat[{Vaish \textit{et al}. \cite{vaish2004using}}]{
\includegraphics[width=2.9cm]{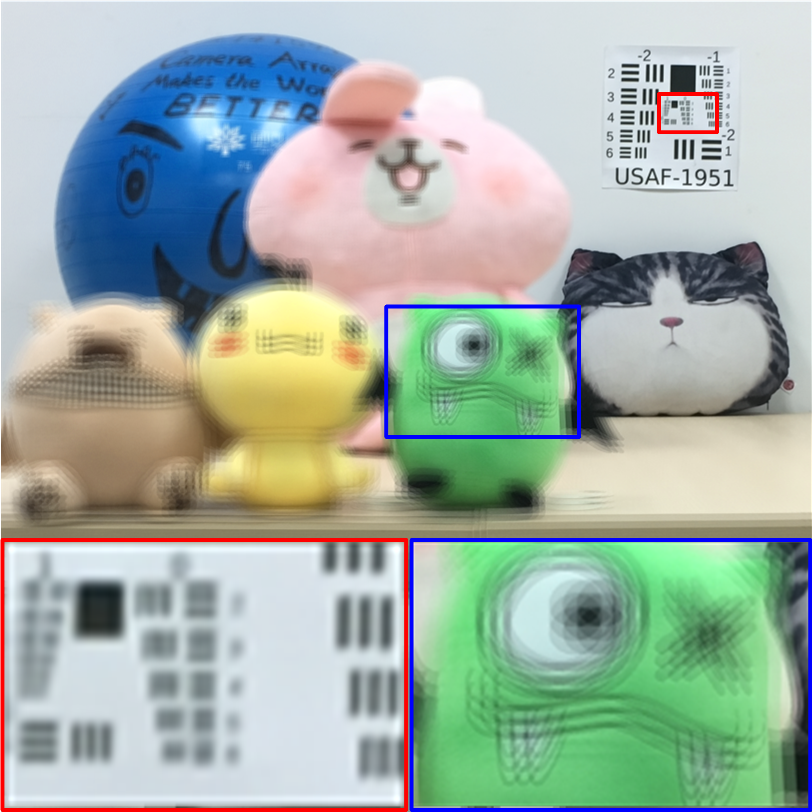}}\subfloat[{Farrugia \textit{et al}. \cite{farrugia2017super}}]{
\includegraphics[width=2.9cm]{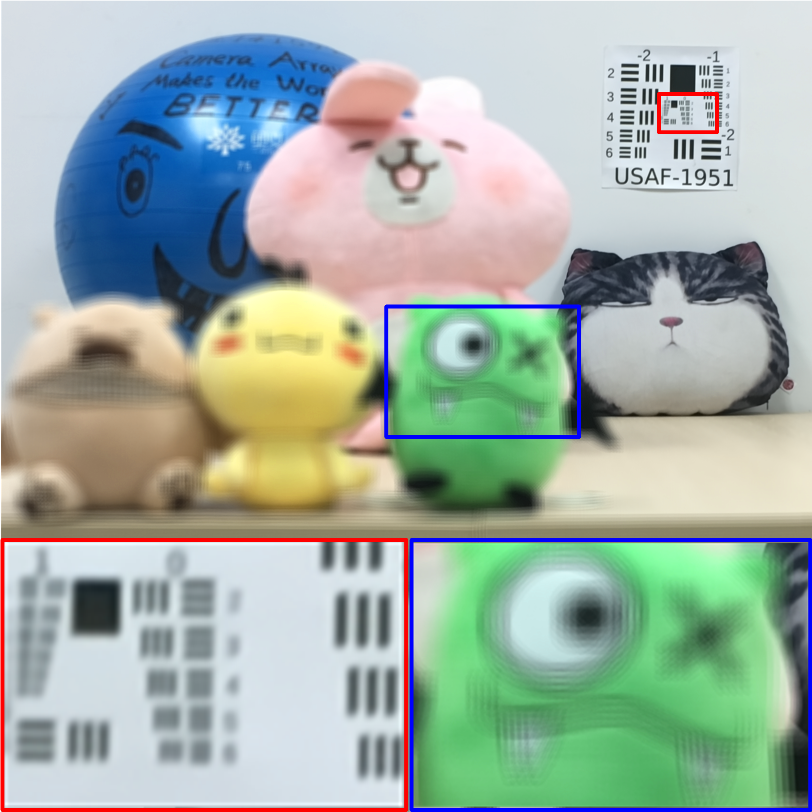}}\subfloat[{Liu \textit{et al}. \cite{liu2016stereo}}]{
\includegraphics[width=2.9cm]{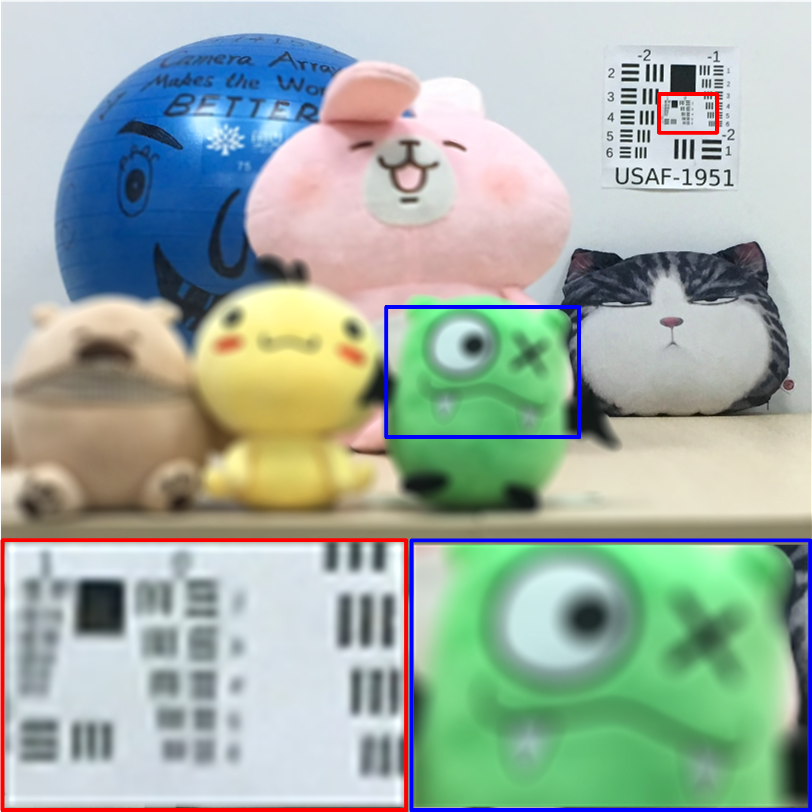}}\subfloat[\textit{Canon} DSLR]{
\includegraphics[width=2.9cm]{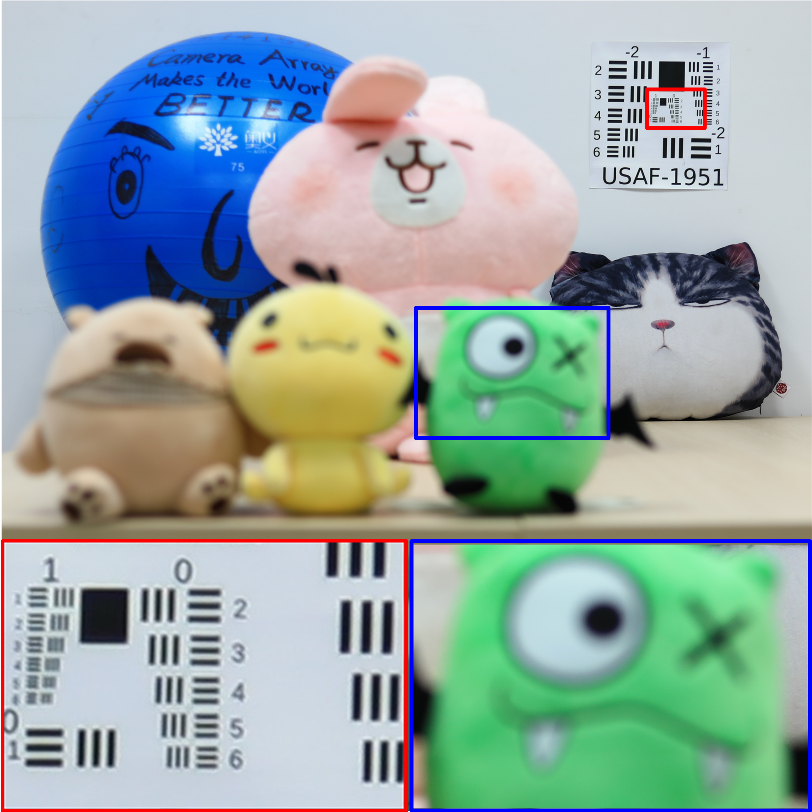}}\subfloat[Ours ($K$=3)]{
\includegraphics[width=2.9cm]{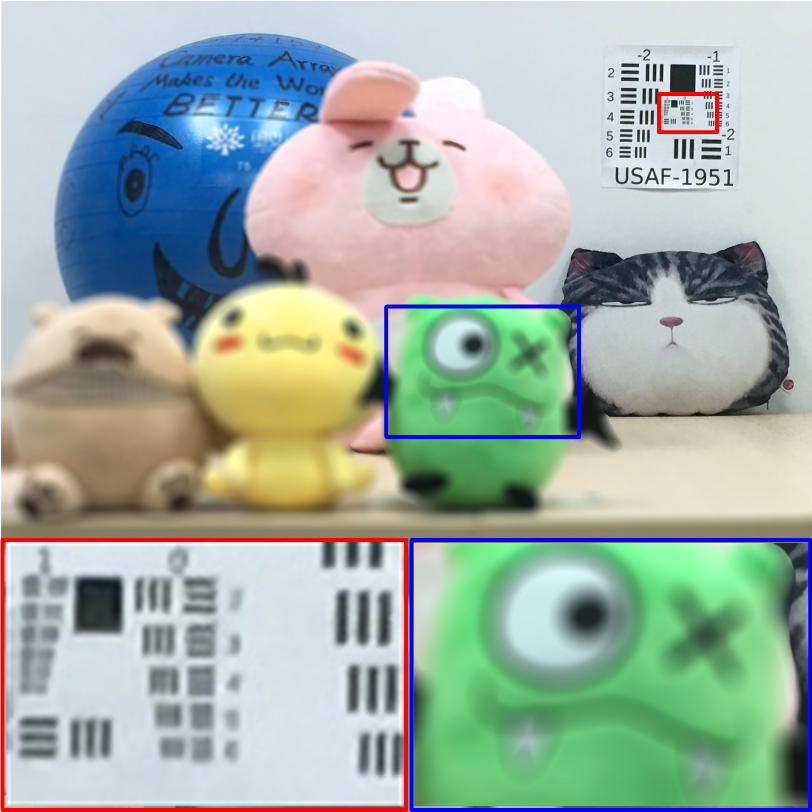}}
\caption{ Results achieved on scene \textit{Dolls} in the self-developed LF dataset. Note that, the
photograph taken by a \textit{Canon} DSLR is used as the \textquotedblleft groundtruth\textquotedblright .}
\vspace{-0cm}
\end{figure*}

From Figs. 3 and 4, we can observe that obvious aliasing artifacts are produced by
\cite{vaish2004using} although shallow DoF is achieved. Farrugia \textit{et al}. \cite{farrugia2017super} improve the image resolution and approximate the real bokeh to a certain extent. That is, the aliasing artifact is alleviated but not eliminated. Liu \textit{et al}. \cite{liu2016stereo} address the aliasing problem but do not improve image resolution, leading to blurring in focused region.
Compared to \cite{vaish2004using}, \cite{farrugia2017super}, and \cite{liu2016stereo}, our method achieves the highest PSNR in focused regions and promising performance in the bokeh. Note that, the bokeh rendered by our method (Fig. 4(f)) is similar to that
generated by the DSLR (Fig. 4(e)). Although the focused region in Fig. 4(f) is not as sharp as that in Fig. 4(e), the natural gap caused by unideal low-cost cameras is partially filled by using our method.

\subsection{Adjustable DoF}
\begin{figure}[t]
\vspace{-0cm}
\centering
\subfloat[$K$=1]{
\includegraphics[width=2.7cm]{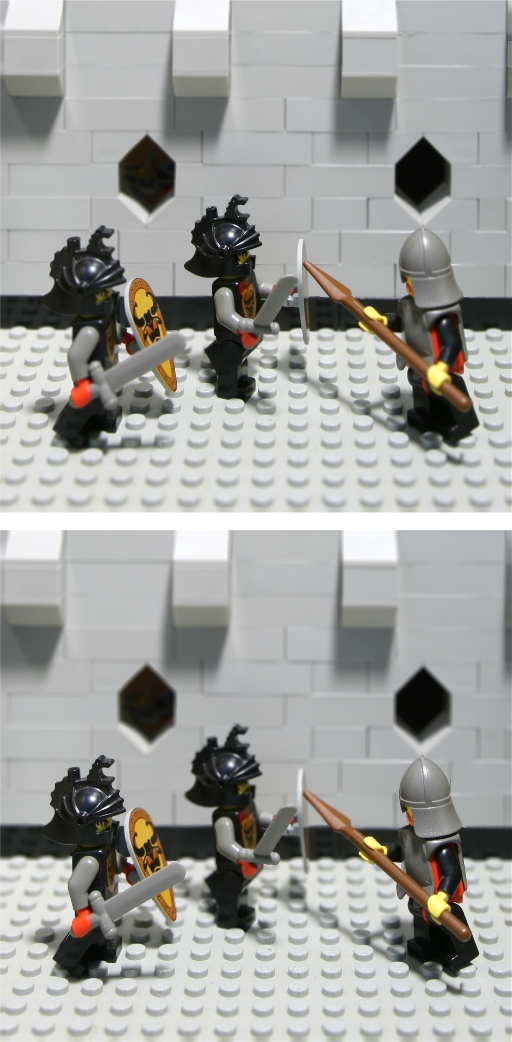}}\subfloat[$K$=2]{
\includegraphics[width=2.7cm]{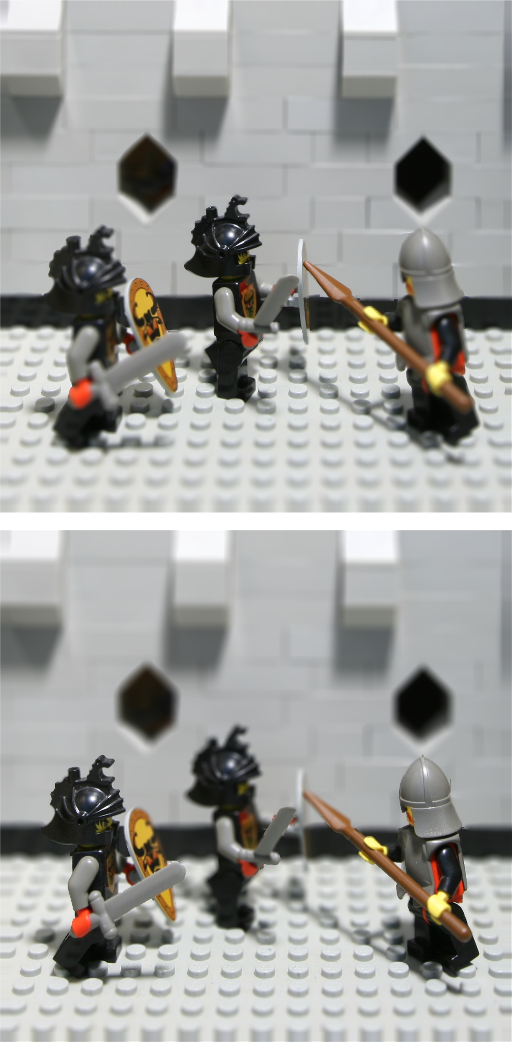}}\subfloat[$K$=3]{
\includegraphics[width=2.7cm]{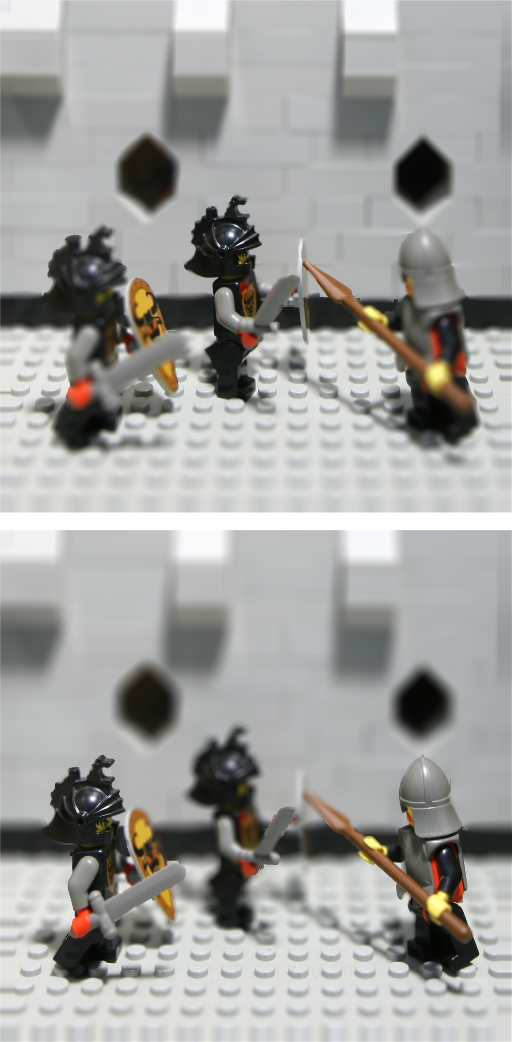}}
\caption{Adjustable DoF achieved by our method on scene \textit{Knights}. Note that, the middle knights are in focus in the top three images
and the side knights are in focus in the bottom three images}
\vspace{-0cm}
\end{figure}
We use our method to refocus images to different depths with different
settings of $K$. As shown in Fig. 5, we can change both the depth in focus and DoF in a post-processing
mode. Note that, DoF post-adjustment is unavailable for DSLRs, traditional refocusing methods
\cite{vaish2004using,vaish2006reconstructing,levoy2004synthetic}, and LF reconstruction methods \cite{farrugia2017super,yoon2017light,Wanner2014Variational}.

\subsection{Influence of Errors in Disparity Estimation}
\begin{figure}[t]
\vspace{-0cm}
\centering
\subfloat[]{
\includegraphics[width=2.1cm]{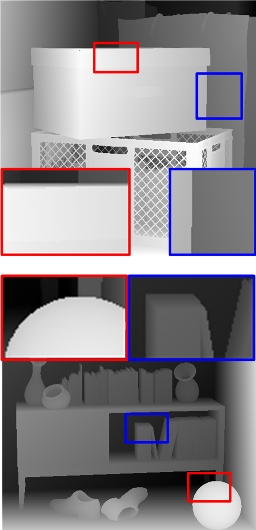}}\subfloat[]{
\includegraphics[width=2.1cm]{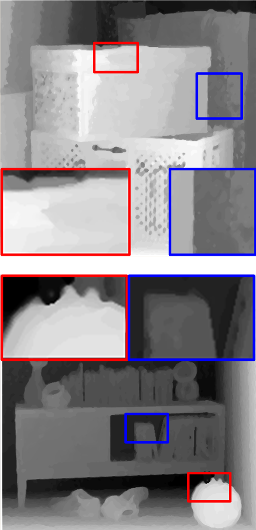}}\subfloat[]{
\includegraphics[width=2.1cm]{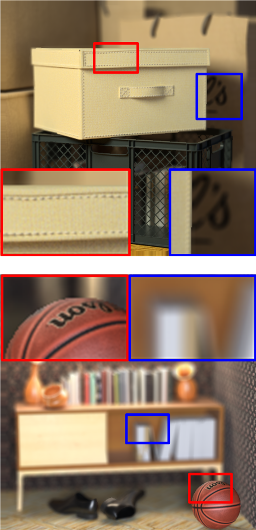}}\subfloat[]{
\includegraphics[width=2.1cm]{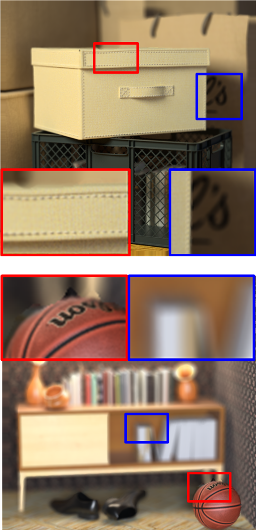}}
\caption{Influence of inaccurate disparity estimation on scene \textit{Box} (top) and \textit{Sideboard} (bottom) in the HCI LF dataset \cite{honauer2016dataset}. (a) Groundtruth disparity, (b) disparity estimated by the method in \cite{wang2018disparity}, (c) results achieved by our method using the groundtruth disparity, (d) results achieved by our method using the estimated disparity in (b).}
\vspace{-0cm}
\end{figure}

We compared the visual performance of our method using the disparity estimated by \cite{wang2018disparity} to that using the groundtruth disparity (provided by the HCI LF dataset \cite{honauer2016dataset}). Results show that the visual performance of our method is partially influenced by the errors in disparity estimation. As shown in Fig. 6, disparity errors on the boundary of the focused region lead to some artifacts. In contrast, our method is robust to minor disparity errors both in bokeh and in focus.

\subsection{Execution Time}

\begin{table}[t]
\scriptsize
\centering
\vspace{-0cm}
\caption{Comparison of Execution Time}
\vspace{-0cm}
\begin{tabular}{|>{\centering}m{2.4cm}|>{\centering}m{1.0cm}|>{\centering}m{1.0cm}|>{\centering}m{0.9cm}|>{\centering}m{1.1cm}|}
\hline
 & Vaish \textit{et al}. \cite{vaish2004using}   & Farrugia \textit{et al}. \cite{farrugia2017super}  & Liu \textit{et al}. \cite{liu2016stereo}  & Ours ($K$=2)\tabularnewline
\hline
\textit{Knights} ($512\times512$) & 1.469 s & 2518 s & 42.59 s & 165.5 s \tabularnewline
\hline
\textit{Cards} ($512\times512$) & 1.446 s & 1732 s & 38.27 s & 174.0 s \tabularnewline
\hline
\textit{Chess} ($700\times400$) & 1.902 s & 2238 s & 48.74 s & 184.0 s \tabularnewline
\hline
\textit{Dolls} ($1024\times1024$) & 6.218 s & 9548 s & 153.5 s & 655.8 s \tabularnewline
\hline
Average of dataset \cite{vaish2008new} & 1.789 s & 2381 s & 46.83 s & 185.2 s \tabularnewline
\hline
\end{tabular}
\vspace{-0cm}
\end{table}

\begin{figure}
\centering
\includegraphics[width=5.5cm]{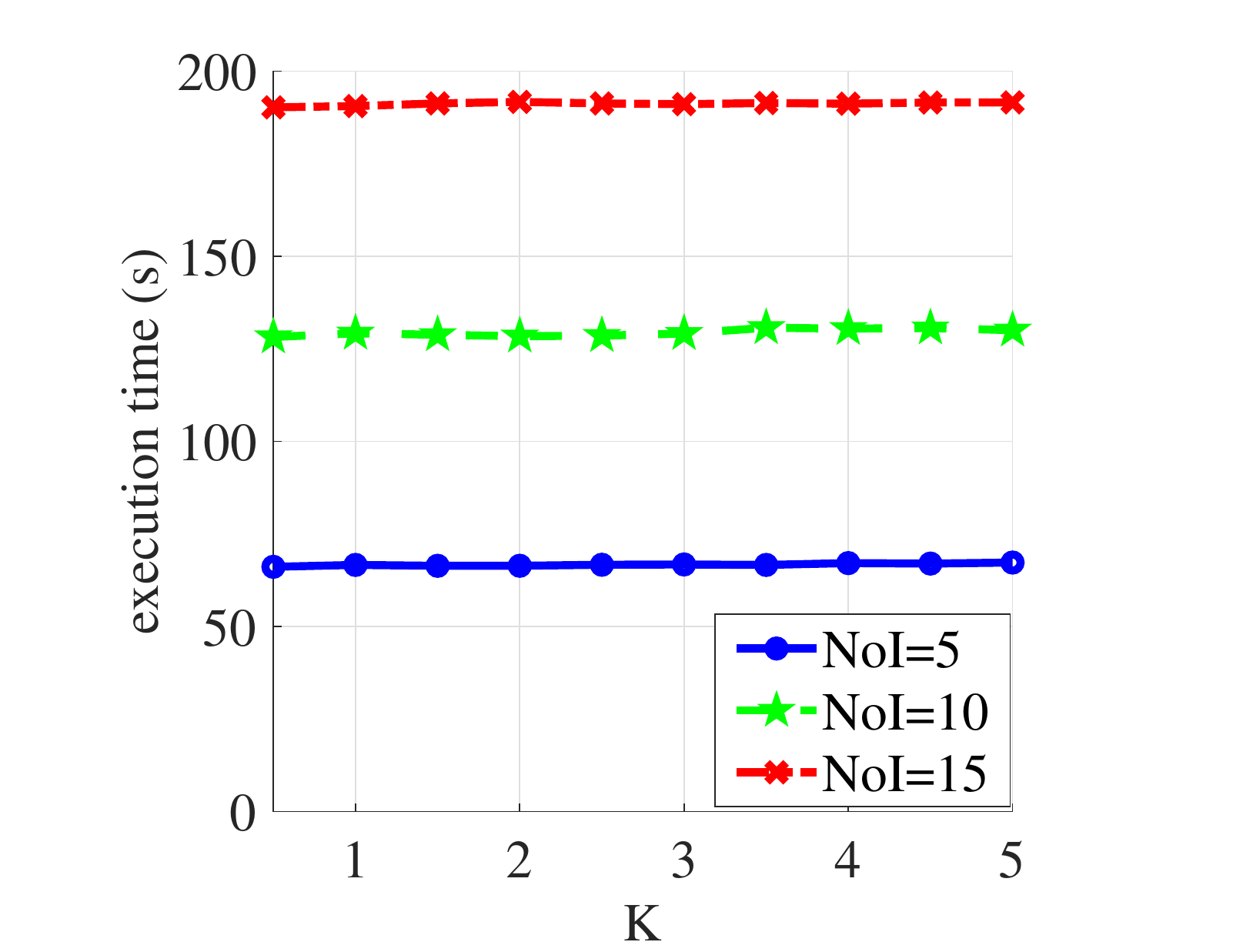}
\vspace{-0cm}
\caption{Execution time of our method achieved on scene \textit{Knights} with different
settings of $K$ and NoI.}
\vspace{-0cm}
\end{figure}

It can be observed from Table II that the shortest running time
is achieved by \cite{vaish2004using} since no depth estimation, pixel
mapping or filtering is required. In contrast, \cite{farrugia2017super} is the most
time-consuming method, and its running time increases dramatically as image resolution increases. Method in \cite{liu2016stereo}
also achieve a relative short execution time because it does not require view rendering or SR, but the resolution
is consequently limited. Since an efficient bokeh rendering scheme is adopted in our method, the execution time of our method is only $5\%$ to $10\%$ of that of \cite{farrugia2017super}.

As the execution time of our method may depend on $K$ (which determines the size of the filter
kernel) and NoI, we further investigate the execution time of our method with
different settings of $K$ and NoI. As shown in Fig. 7,
NoI has much more influence on the overall execution time
than $K$ because the adopted bokeh rendering approach is computationally
efficient. Since a too large bokeh (i.e., too shallow DoF) will destroy
the esthetic value of a photograph, we need to determine the DoF
empirically based on the scenario. Generally, a shallow DoF (e.g., $K$=3) is preferred for close-shot photographs (e.g., scene \textit{Knights}), and a deep DoF (e.g., $K$=1) is suggested for distant scenes (e.g., landscape photograph).

\section{Conclusion}
In this paper, we propose a method to refocus images captured by a camera array. We formulate an SR model and design
an anisotropic filter to increase the image resolution and render
the bokeh. The superiority of our method is demonstrated by
experiments on both public and self-developed datasets. Our
method enables camera arrays to produce aesthetical photographs with acceptable computational cost.

\bibliographystyle{IEEEtran}
\bibliography{SPL}

\end{document}